\documentstyle[12pt]{article}

\newcommand{\be}{\begin{equation}}
\newcommand{\ee}{\end{equation}}
\newcommand{\bea}{\begin{eqnarray}}
\newcommand{\eea}{\end{eqnarray}}

\begin{document}
\begin{titlepage}

%\flushright{To Appear: }

\vspace{1in}

\begin{center}
\Large
{\bf COLLIDING STRING WAVES AND DUALITY }

\vspace{1in}

\normalsize

\large{ Ashok Das$^a$, J. Maharana$^b$  and  A. Melikyan$^a$}

\normalsize
\vspace{.5in}

$^a${\em Department of Physics and Astronomy \\
University of Rochester, NY 14627-0171, USA \\}

\bigskip
$^b$ {\em Max-Planck-Institut f\"ur Gravitationsphysik \\
Albert-Einstein-Institut \\
14476 Golm, Germany \\

\bigskip
and\\
\bigskip
Institute of Physics, \\
Bhubaneswar - 751005, India \\}

\end{center}

\vspace{1in}

\baselineskip=24pt
\begin{abstract}
The collision of plane waves corresponding to massless states of closed
string is considered in $D$-dimensional space-time.  The reduced tree level
effective action is known to be manifestly $O(d,d)$ invariant, $d$ being
the number of transverse spatial dimensions in the collision process. We
adopt a coset space reformulation of the effective two dimensional
theory and discuss the relation of this process with classical integrable
systems in two dimensions in the presence of gravity. We show how
it is possible to generate new backgrounds for the scattering process,
from known background solutions to the equations of motion,  in the
coset reformulation. We present explicit calculations for the case of
four space-time dimensions as an illustrative example.

\end{abstract}

\vspace{.7in}
 
\end{titlepage}

%\section{INTRODUCTION }

%\setcounter{equation}{0}

%\def\theequation{\thesection.\arabic{equation}}

Recently, there have been attempts to study collision of plane waves in the
frame work of string theory \cite{fkv,bv}. A scenario is envisaged where 
infinite plane-fronted waves undergo a head on collision. One of the  goals
of the  investigation of  this problem is to gain insight into the
mechanism for the creation of the Universe in the pre-big bang (PBB)
 hypothesis \cite{pbb, rev}.
In this scenario, it is proposed that the Universe, in its infancy, is
in the weak coupling 
regime with low curvature, where we can trust the low energy string effective
action. In this phase, we can admit the Minkowski vacuum order by order
in the perturbative frame work. The Universe, in this early epoch, can be
described as a superposition of massless waves, travelling in all directions,
whose energies lie below the string scale. 
These
waves  collide in the process of propagation, constructively interfere 
and eventually
gravitational collapse takes place, when over-dense regions are
formed. It has been argued by Buonanno, 
Damour and Veneziano \cite{bdv}   that the interior of an adequately large
and collapsing region may give birth to a Universe which possesses 
attributes of our own Universe. In other words, collisions of plane waves 
could result in nontrivial geometries in the interaction region of the
waves, leading to the creation of a Universe as proposed in the PBB
scenario. The 
singularity, arising from the interaction process of the plane waves,
finds an interpretation as a cosmological singularity from the
perspective of the PBB proposal. In the study of the scattering process, when
one solves the underlying equations in the interaction region, the 
resulting metric assumes Kasner's form. Moreover, the exponents fulfill
the requirements of the PBB hypothesis in the space of admissible parameters.
Furthermore, it has been argued that initial states of the PBB cosmology
could be thought of as the plane waves associated with the massless spectrum
of strings \cite{reza}. It is important to note that Kasner solutions have
attracted considerable attention from various perspectives \cite{k1,k2}. 

It is  ideal to consider configurations of plane 
waves of infinite front  undergoing collisions and several interesting
results have already emerged in recent studies \cite{fkv,bv}. In the
work of FKV  
\cite{fkv}, the collision process was studied in four ($D=4$) dimensional 
space-time for graviton and dilatonic waves. Subsequently, 
Bozza and Veneziano \cite{bv} examined the case of arbitrary $D$-dimensional
space-time (with $D-2=d $ transverse dimensions) and incorporated the presence
of the antisymmetric tensor background through implementation of the $O(d,d)$
symmetry, since the effective two dimensional tree level action is invariant
under this target space symmetry. We mention in passing that collision
of plane waves has been studied in detail in general theory of relativity 
\cite{book}; 
however, interest, in this, has been revived since there is an
interesting connection 
between this process and the issue of the creation of the Universe in
the PBB scenario. Since the problem under consideration is described by
a two dimensional effective action, it is natural to expect 
 an intimate connection with integrable systems \cite{das}. Indeed, there has
 been a 
considerable amount of work in investigating a class of two dimensional
field theories in curved space-time and super-gravity theories to 
study their integrability properties
\cite{nico1}.

It is expected that the techniques adopted in the study of integrable systems
 will unravel other fascinating features of 
 the problem under consideration. In the past, the
symmetries of two dimensional string effective action have been investigated
in some detail by several authors \cite{jm,john,sen,others}. Moreover,
if we adopt the string frame metric in expressing the reduced action, then 
the manifest $O(d,d)$ symmetry can be exploited to generate new background
configurations from  known solutions of the  equations of motion.

The purpose of this note is two fold. First, we shall argue how the effective
action, studied from a different perspective, can be shown to have a
direct  connection with integrable systems. 
This is achieved rather elegantly by adopting
the coset space reformulation of the action as was derived by Schwarz
and one of the authors \cite{jmj}. Thus the action can be expressed in a
current-current form involving the scalar fields (moduli). 
 Our
second result is the special case of collision of waves in four space-time
dimensions (i.e., transverse dimensions $d=2$). In this case, we start
with  a background where the
antisymmetric tensor field is vanishing along transverse directions
while the
metric and the dilaton are nonzero. We generate a nontrivial $B$-field
along the transverse directions by exploiting a property of the coset 
reformulation, so that the new background configuration can be extracted
rather easily. This technique is applicable to the case with arbitrary 
$d$ transverse dimensions and is a property of the reformulation alluded to
above.

The $D$-dimensional tree level string effective action is
\bea
S_D=\int d^Dx{\sqrt {-\cal G}}e^{-\phi} \left(R_{\cal G} +(\partial \phi)^2
-{1\over {12}}H_{\mu\nu\rho}H^{\mu\nu\rho}\right) \eea
Here ${\cal G}_{\mu\nu}$ is the $D$-dimensional metric in the string frame, 
$\cal G$ is its 
determinant and $R_{\cal G}$ is the curvature scalar derived from this
metric. Note that $\phi$ is
the dilaton and $H_{\mu\nu\rho} =\partial _{\mu}B_{\nu\rho}+\partial _{\rho}
B_{\mu\nu}+\partial _{\nu} B_{\rho\mu}$ is the field strength
associated with the anti-symmetric field $B_{\mu\nu}$. When one 
studies the problem of head on collision of two plane-front waves of infinite
extension, it is possible to define a coordinate
 system where the waves travel along
$x^1$ direction. If we assume that there is translational symmetry in the
transverse directions, then there will be Abelian isometries along these,
$D-2=d$, directions. Thus the process is described by a two dimensional
reduced effective effective action, which can be derived following
the prescription of ref. \cite{jmj}
 \bea
\label{twod}
S=\int dx^0dx^1{\sqrt {-g}}e^{-\bar {\phi}}\left(R+(\partial {\bar \phi})^2+
{1\over 8}{\rm Tr}(\partial _{\alpha}M^{-1}\partial ^{\alpha}M)
\right) \eea

In general, there will be additional terms in the reduced action
corresponding to $d$ Abelian gauge fields from the original metric and another
set of $d$ gauge fields from the antisymmetric tensor $B_{\mu\nu}$ as a
result of dimensional reduction \cite{jmj}.
 Furthermore, there would have been the field
strength of the two dimensional tensor field $B_{\alpha\beta}$ where
$\alpha , \beta =
0,1 $ are spacetime indices. Since we are
effectively in two space-time dimensions, we have dropped the gauge field terms
and, in the same spirit, have not kept field strengths of $B_{\alpha\beta}$,
which can be removed if it depends only on coordinates $x^0 $ and $x^1$.
 The other
terms appearing in the above action are defined as follows: 
$g_{\alpha \beta}$ with $\alpha ,\beta =0,1$ 
is the two dimensional space-time metric, and $R$ is the corresponding
curvature scalar.
The shifted dilaton is  ${\bar \phi}=
\phi -{1\over 2} {\rm log}~{\rm det}~G $,  with $G$  the
metric corresponding to the transverse coordinates $x^i, i=2,3.., D-1$. $M$ is
a $2d \times 2d$ symmetric matrix  
\be
\label{mmatrix}
M = \pmatrix {G^{-1} & -G^{-1} B\cr
BG^{-1} & G - BG^{-1} B\cr} \ee
where $B$ represents the moduli coming from the reduction of the
$B$-field  in $D$ space-time
dimensions. The symmetric nature of $M$ is evident since $G$ is
symmetric and $B$  is antisymmetric. 

The reduced effective action is
invariant  under global $O(d,d)$ transformations
\be g_{\alpha\beta} \rightarrow g_{\alpha\beta}, ~~ 
 {\bar \phi} \rightarrow {\bar \phi} \ee
\be M \rightarrow \Omega ^TM\Omega \ee
where $\Omega \in O(d,d)$ is the transformation matrix. The $O(d,d)$ metric
\be \eta=\pmatrix {0 & {\bf 1}\cr {\bf 1} & 0\cr} \ee
remains invariant under the non-compact global transformations, $\bf 1$ being
the $d$-dimensional unit matrix. Furthermore, it is easy to check that the
 matrix $M$
belongs to $O(d,d)$. 

The relevant  field equations are
\be
\label{dil}
R+2g^{\alpha\beta}D_{\alpha}D_{\beta} {\bar \phi}-g^{\alpha\beta}
\partial_ {\alpha}{\bar \phi}\partial_ {\beta}{\bar \phi}+{1\over 8}g^{\alpha
\beta}{\rm Tr}(\partial _{\alpha}M^{-1}\partial _{\beta}M)=0 \ee
\be
R_{\alpha\beta} +D_{\alpha}D_{\beta} {\bar \phi} +{1\over 8} {\rm Tr}
(\partial _{\alpha}M^{-1}\partial _{\beta} M)=0 \ee
which result from the variation of the effective action
with respect to the shifted dilaton and the metric $g_{\alpha\beta}$
respectively. The equation 

\be D^{\alpha}D_{\alpha} e^{-\bar \phi}=0 \ee
follows from the above two equations in a straightforward manner.
The equation of motion for $M_{ij}$ is 
\be
\label{matrix}
 \partial _{\alpha}(e^{-\bar \phi}{\sqrt {-g}}g^{\alpha\beta} M^{-1}
\partial _{\beta}M)=0 \ee
which is obtained by a constrained variation of the effective action,
taking account of the fact that $M \in O(d,d)$. Equations 
(\ref{dil})-(\ref{matrix}) can be simplified further in the light cone
coordinates, if we recall that
the two dimensional metric can be written as the flat metric times a
conformal factor, i.e.,
$ g_{\alpha\beta}=e^F\eta _{\alpha\beta}$.  
In this case, the relevant equations take a rather simple form as has
been shown in
\cite{bv}
\be \partial _u\partial _ve^{-\bar \phi}=0 \ee 
which is also equivalent to $\partial _u\partial _v{\bar \phi}=\partial _u{\bar
\phi}\partial _v{\bar \phi} $ and
\be \partial _u(e^{-\bar \phi}M^{-1}\partial _vM)+\partial _v(e^{-\bar \phi}
M^{-1}\partial _uM)=0 \ee
\be \partial_u^2{\bar \phi}- \partial _uF\partial _u{\bar \phi}+{1\over 8}
{\rm Tr}(\partial _uM^{-1}\partial _uM)=0,~ {\rm another~ equation~  with ~}
u\rightarrow v \ee
\be \partial _u\partial _v{\bar \phi}-\partial _u\partial _vF+{1\over 8} 
{\rm Tr}(\partial _uM^{-1}\partial _v M)=0 \ee
where $u={1\over {\sqrt 2}}(x^0-x^1) ~{\rm and }~ 
\,v={1\over {\sqrt 2}}(x^0+x^1)$.

Let us recapitulate very briefly how solutions are obtained for
the scattering of the plane waves. The waves collide at $u=0 ~{\rm and}~
v=0$ 
which corresponds to $x^0=x^1=0$. The fronts of the two waves are defined
to be at $u=0$ and $v=0$ respectively. In order to describe the
collision  process, it is
convenient to divide space-time into four regions. 
First, one considers a region which
is in front of the plane waves, called region $I$ such that $u,v <0$
and this is before the collision occurs. The metric is Minkowskian  
here and the string coupling is small, so that we are in the
perturbative  regime.
Here, for the dilaton we have $\phi =\phi _0$, with $e^{\phi _0} <<1$
and we  can fix 
$\bar \phi=0$; and  take $B=0$. The line element is
\be ds^2_I=-2 dudv+ dx^idx_i \ee
in this region and we adopt the convention of summation over repeated
indices throughout. Before
the collision occurs, a wave is coming from the left, called region
$II$, where $u>0, v<0$ and the backgrounds $G^{II}_{ij}(u), B^{II}_{ij}(u)
~{\rm and }~\phi ^{II}(u)$ are only $u$-dependent and the superscript
specifies that they are in region $II$. The field equations enable  us to 
choose the conformal factor, $F=0$ in region $II$ and the line element has the
 following form
\be
ds^2_{II}= -2dudv+G^{II}_{ij}(u)\,dx^idx^j \ee
The other wave front is coming from the right, i.e.,
$v>0 ~{\rm and} ~u<0$, which is denoted by $III$. Here the backgrounds
depend  only
on $v$ and they are  designated with the superscript $III$. We shall not 
display the line element explicitly;  it has the same form as that for $II$, 
except that $G^{II}_{ij}(u)$ is replaced by
$G^{III}_{ij}(v)$. Finally, we have
region $IV$, where interaction takes place and the conformal factor, $F(u,v)$,
makes its appearance in equations and the backgrounds $G^{IV}_{ij},
B^{IV}_{ij} ~{\rm and }~ \phi ^{IV}$ depend on both $u~{\rm and}~ v$
so that the line element takes the form
\be ds^2_{IV}=-2e^{F(u,v)}dudv+G^{IV}_{ij}(u,v)\,dx^idx^j \ee
The solutions to background equations of motion for this system in
the string theory have been discussed in  \cite{fkv,bv}, especially
the solution for the diagonal form of $G$ has been dealt with in detail.
Our goal, in the first step, is to show the connection between this
problem and integrable systems \cite{das}. In order to achieve this, first
we have to cast the action in a different form.

We may recall that the reduced effective action admits a coset space
reformulation as was demonstrated in ref. \cite{jmj}. One introduces
a triangular matrix form for $M$, such that
$ M=VV^T$  and
\be 
V = \pmatrix {E^{-1} & 0 \cr
BE^{-1} & E^T \cr} \ee
 where $B$ is the antisymmetric tensor $B_{ij}$ and $E$ is a vielbein such
that  $(E^TE)_{ij}=G_{ij}$. Also it is easy to show that $V, V^{T} \in
O(d,d)$. It
was shown that one can introduce an arbitrary matrix $V$ which belongs to
$O(d,d)$ (not the special form given above) and write an action which
is invariant under global $O(d,d)$ transformation as well as 
under local gauge transformations of its maximal compact subgroup $O(d)
\times O(d)$. Then, one can make a gauge choice to obtain the  form of the 
reduced action that appears in standard dimensional reduction with manifest
$O(d,d)$ symmetry. Of course, for the problem at hand, we shall adopt the
results of \cite{jmj} accordingly. First, we introduce a transformation
to go over to the diagonal form of $O(d,d)$ metric,  $\sigma =
\pmatrix {{\bf 1} & 0 \cr 0 & -{\bf 1}\cr} $, 
 $\bf 1$ being  the $d \times d$ unit matrix, through the relation
\be L^T \eta L=\sigma ,~~~~L={1\over {{\sqrt 2}}}\pmatrix{{\bf 1} & {\bf -1}\cr
{\bf 1} & {\bf 1}\cr } \ee
The $M$ and the $V$ matrices transform as follows:
\be M\rightarrow {\tilde M}= L^TML,~~~ V\rightarrow {\tilde V}= L^TVL \ee

Let us next define a current
\be 
\label{PQ}
{\tilde V}^{-1}\partial _{\alpha}{\tilde V}=P_{\alpha}+Q_{\alpha} \ee
where $\tilde{V}^{-1}\partial_{\alpha} \tilde{V} \in O(d,d)$ and
$Q_{\alpha} \in O(d)\times O(d)$, the maximal
compact subgroup. It is  straightforward  to show that
\be 
\label{eqsigma}
{\rm Tr}(\partial _{\alpha}M^{-1}\partial _{\beta} M) = -4\, {\rm Tr}\,
P_{\alpha}P_{\beta} \ee
It follows from the symmetric space automorphism property of the coset
${O(d,d)\over {{O(d)\times O(d)}}}$ that $P_{\alpha} ^T =P_{\alpha}$, which
belongs to the complement and $Q_{\alpha} ^T=-Q_{\alpha}$ as
$Q_{\alpha} \in O(d)\times O(d)$. 
Thus, we can write $P_{\alpha}={1\over 2}[{\tilde V}^{-1}\partial
_{\alpha}{\tilde V} 
+({\tilde V}^{-1}\partial _{\alpha}{\tilde V})^T]$ and then
(\ref{eqsigma})  follows
in a straightforward manner. We
mention another important point in passing, namely, 
 that, under an  $O(d) \times O(d)$ gauge transformation ${\tilde V}
\rightarrow {\tilde V} h(x), h(x) \in O(d) \times O(d) $, 
 $P_{\alpha} ~{\rm and}~ Q_{\alpha} $ transform 
as  follows 

\be P_{\alpha} \rightarrow h(x)^{-1}P_{\alpha}h(x),~~~
Q_{\alpha}  \rightarrow
h^{-1}(x)Q_{\alpha}h(x)+h^{-1}\partial _{\alpha}h(x) \ee
and this transformation leaves (\ref{eqsigma}) invariant.
Notice the form of (\ref{eqsigma}): in flat space-time, this describes
a $\sigma$-model in two dimensions. Furthermore, it is possible to
introduce a constant spectral parameter and a zero curvature condition
by taking 
a suitable combination of $P_{\alpha} ~{\rm and}~ Q_{\alpha}$ to show the
integrability properties of the classical theory following well known
procedures \cite{das}. For the case at hand, the full action consists
of kinetic energy terms for the conformal factor $F$ and the shifted
dilaton $\bar \phi$ in addition to the piece given by (\ref{eqsigma}).  
 One can scale the metric $g_{\alpha\beta}
\rightarrow e^{\bar \phi} g_{\alpha\beta}$ to get rid of the kinetic
term of $\bar \phi$ in action (\ref{twod}). Then, for the modified
action, one can follow the procedure described in 
\cite{nico1} to establish relation
of this process with classical integrable systems.
Let us consider the special case, as an illustrative example, when the metric
$G_{ij}$ is diagonal as was the case in \cite{fkv,bv} and $B_{ij}=0$. The
corresponding $V$-matrix is symmetric and so is $\tilde V$. Thus it
follows from 
(\ref{PQ}) that $Q_{\alpha}=0$ and one needs to suitably reformulate 
integrability conditions for this case. When one studies integrability
properties of two dimensional models in curved space-time, the spectral 
parameter assumes space-time dependence. Furthermore, ${\tilde V}(x^0,x^1)
\rightarrow
{\hat V}(x^0,x^1,t)$ such that
\begin{eqnarray} 
\label{linears}
{\hat V}^{-1}\partial _v{\hat V} & = & {{1- t}\over {1+
      t}}P_v \nonumber\\
{\hat V}^{-1}\partial _u {\hat V} & = & {{1+t}\over {1-t}}P_u 
\end{eqnarray}

The equations of motion and integrability condition

\bea
\partial _{\alpha}({\hat V}^{-1}\partial _{\beta}{\hat V})-\partial _{\beta}(
{\hat V}^{-1}\partial _{\alpha}{\hat V})+[{\hat V}^{-1}\partial _{\alpha}{\hat
V} , {\hat V}^{-1} \partial _{\beta}{\hat V}] =0 \eea
are compatible provided the spectral parameter satisfies 
\be \partial _{\alpha } t=-{1\over 2} \epsilon _{\alpha \beta}\partial ^{\beta}
(e^{-\bar \phi}(t+{1\over t})) \ee
The solution to the above first order equation is
\be {{1-t}\over {1+t}}=\sqrt { {w-e^{-{\bar \phi}(u)}}\over {{w+e^{-{\bar \phi}
(v)}} }} \ee
where $w$ is an integration constant. We shall see that, in region
$IV$,  $e^{- \bar \phi}$ is expressed as a sum of two solutions, one
depending  on $u$ and the other on $v$. The next step is to solve for
the  monodromy matrix. It is
well known that the linear system, defined through (\ref{linears}), is
invariant under the generalization of symmetric space automorphism. One defines
\be \tau ^{\infty} {\hat V}(t)=({\hat V} ^T)^{-1}({1\over t}) \ee
In terms of Lie algebra elements, this implies $P_{\alpha}\rightarrow -
P_{\alpha}, ~~ t\rightarrow {1\over t}$. Furthermore, the current
${\hat V}^{-1}\partial _{\alpha}{\hat V}$ is invariant under $\tau ^{\infty}$
transformation. The monodromy matrix 
\be {\cal M}={\hat V}\tau ^{\infty}({\hat V})^{-1} = {\hat V}(x,t){\hat
V}^T (x, { 1\over t}) \ee
plays a fundamental role in the reconstruction of solutions as well as
in the study of integrability properties from this point of view. We
shall present the explicit form of $\cal M$ for the four dimensional
case,  when
we discuss solutions.

Another aspect of this reformulation is that 
 one can generate new background configurations from known solutions 
as follows:
First solve for equations of motion for a given background. Then implement
an $O(d,d)$ transformation on $\tilde V$. In general, the resulting
$V$ obtained from transformed $\tilde V$ will not maintain its triangular 
form, which is quite essential to get the desired form of the $M$-matrix.
At this stage, one can further make a transformation $h\in O(d)\times
 O(d)$  to bring
the transformed $V$, call it $V'$, to a triangular form. 

Let us consider, to be specific, the scattering of plane fronted waves
in $D=4$ for which the duality symmetry is $O(2,2)$. Thus $G_{ij} 
~{\rm and}~ B_{ij}$ are $2 \times 2$ matrices. If we choose a background
configuration where $G_{ij}$ is diagonal and
 $B_{ij}=0$, then 
\be  V=\pmatrix{E^{-1} & 0\cr  
            0 & E^{T} \cr} \ee
Here, $E$ is the corresponding zweibein with a choice $E=\pmatrix {E_1 & 0 \cr
0 & E_2 \cr }$  and that $G=E^TE$. We recall that  under a
global $O(2,2)$ transformation, $V \rightarrow \Omega ^TV$ with
$\Omega \in O(2,2)$. 
Let us choose
\be \Omega =\pmatrix {a{\bf 1} & b\epsilon \cr
                     -c\epsilon & d{\bf 1} \cr} \ee
where ${\bf 1}$ is unit $2\times 2$ matrix and $ \epsilon$ is the antisymmetric 
 $2\times 2$  matrix. Note that $a,b,c~{\rm and }~  d$ are real
numbers satisfying $ad-bc=1$,  in order that the $4\times 4$ matrix 
$\Omega \in O(2,2)$. Under such a transformation, 
\be  V \rightarrow \Omega ^T V=
\pmatrix{aE^{-1} & c\epsilon E^T \cr
                              -b\epsilon E^{-1} & dE^T \cr} \ee
Thus, as mentioned, the transformed $V$ is not in the desired triangular
 form, although we can read off the new transformed backgrounds to be
\be E'={E\over a}, ~{\rm}~ B'=-{b\over c} \epsilon \ee
Note that the resulting $B'$ is just a constant. 

However, let us recall that we also have the freedom of making 
a local $O(2) \times O(2)$ transformation under which,
${\tilde V} \rightarrow {\tilde V}h(x), ~h(x) \in O(2) \times O(2)$. 
 Let us choose the $4\times 4$
matrix $h$ to be space-time
 independent and of the form
\be h=\pmatrix { \gamma{\bf 1} & \delta \epsilon \cr
                \delta \epsilon & \gamma {\bf 1} \cr } \ee
It is easy to check that, $hh^T=h^Th ={\bf 1}$ provided, $\gamma
^2+\delta ^2=1$. 
The $O(2)\times O(2)$ nature of the matrix becomes more transparent,
if we go  to the diagonal basis for the metric
of $O(2,2)$ where we see that  
$ h\rightarrow L^ThL = {\rm diag}~ (\gamma {\bf 1}+\delta
\epsilon ,  \gamma {\bf 1}-\delta \epsilon) $ and the off diagonal elements are
zero. 
After $V$ gets transformed under both $\Omega ~{\rm and }~   h$,
 we finally get
\be V \rightarrow \Omega ^TVh=\pmatrix {a\gamma E^{-1}+c\delta E^T\epsilon &
a\delta E^{-1}\epsilon +c\gamma \epsilon E^T \cr
-b\gamma\epsilon E^{-1}+d\delta E^T\epsilon & 
-b\delta \epsilon E^{-1}\epsilon +
d\gamma E^T \cr } \ee
If this transformed $V$ is to be of triangular form, then its 
\lq\lq $12$'' component is to 
be set to zero,
 $a\delta E^{-1}\epsilon+c\gamma \epsilon E^T =0$ 
which determines, $\gamma ~{\rm and }~ \delta$ to be
\be \gamma={a\over {{\sqrt {a^2+c^2E^2_1E^2_2}}}}, ~~ 
\delta=-{{cE_1E_2}\over {{\sqrt
{a^2+c^2E^2_1E^2_2}}}} \ee 
and they automatically satisfy the requirement $\gamma ^2+\delta ^2 =1$.
The new backgrounds, denoted by primes (not to be confused with derivatives) are
\be  G'=(a^2G^{-1}-c^2\epsilon G\epsilon )^{-1} \ee
\be 
\label{eqb}
 B'=-{{ab+cd E^2_1E^2_2}\over {a^2+c^2E^2_1E^2_2}}\epsilon \ee
and we note that the new moduli $ B' _{ij}$ are space-time dependent.
Furthermore, $ G_{ij}'$ depends on $G ~{\rm and }~ G^{-1}$ and 
is diagonal. For this choice of $\Omega ~{\rm and} ~ h$ we have generated
nontrivial
backgrounds. 

Let us now discuss some of the essential properties of the solutions
 in  region $IV$, since this is the 
 region where collision occurs. When we implement an $O(d,d)$ transformation,
the shifted dilaton $\bar \phi$ remains invariant. The solutions to the
equation of motion, for diagonal metric and $B=0$, has been discussed
in \cite{fkv,bv}, where they appear in form of integrals. These results
are generalization of the case of pure gravity considered by Szekeres \cite{sz}
and subsequently by Yurtsever \cite{yr}. The asymptotic behavior of 
the solutions are of interest, since one of the aims is to examine whether
a singularity appears in the backgrounds. Let us express \cite{fkv} the diagonal
elements $G_{ij}, i,j =2,3$, corresponding to directions $x^2$ and $x^3$, in 
the following form $G_{22}={\rm exp}(\lambda +\psi)$ and $G_{33}={\rm exp}
(\lambda -\psi)$. Thus $\lambda ={1\over 2}{\rm log}~{\rm det}~G$. 
The dilaton, in region $IV$, can be written as
\be 
\label{fourd}
e^{-{\bar \phi}(u,v)}=e^{-{\bar \phi}_{II}(u)}+e^{-{\bar \phi}_{III}(v)}-1
\ee
That it is a sum of a function of $u$ and of $v$ in region $IV$, follows from
the equation of motion for $\bar \phi$ as well as the fact 
that the solutions should match smoothly across the boundaries
of the different regions. It is convenient to introduce a new set of variables
\be
r=2\,{\rm exp}(-{\bar \phi}_{III}(v))-1, ~~s=2\,{\rm exp}(-{\bar
 \phi}_{II}
(u))-1 \ee  
Thus $r$ and $s$ are functions of the variables $v$ and $u$ respectively.
Let us define a set of coordinates 
\be
\xi={(r+s)\over 2} = {\rm exp}(-{\bar \phi}(u,v)), ~~
 z={(r-s)\over 2} = e^{-{\bar \phi}_{II}
(u)}-e^{-{\bar \phi}_{III}(v)} \ee
Obviously, ${\bar \phi}(u,v)$ is defined only in region $IV$. From now
on, we  are
going to concentrate on the  backgrounds only in region $IV$ and
 therefore, we  suppress the
subscript $IV$ from all of them. Under the change of variables, the conformal
conformal factor appearing in the space-time metric gets affected. The
line element in region $IV$ takes the form (subscript not displayed)
\be ds^2=-e^fd\xi ^2+e^fdz ^2 +G_{22}(dx^2)^2+G_{33}(dx^3)^2 \ee
where $f =f(r,s)$ is the conformal factor and
\be {\bar \phi}= -{1\over 2}{\rm log}~(r+s)=-{\rm log}~\xi \ee
The asymptotic behavior of the functions $\lambda$, $\psi$, appearing in 
$G_{ij}$, the conformal factor, $f$ and dilaton $\phi$ are given by
\be \lambda \sim {\kappa (z)}\,{\rm log}~\xi,~~\psi \sim A(z)\, {\rm
log}~\xi \ee 
\be f \sim B(z)\, {\rm log}~\xi,~~\phi\sim \kappa (z)\,{\rm log}~\xi -1 \ee
where
\bea
\kappa (z)&=&{1\over {\pi\sqrt{1+z}}}\int _z ^1ds[(1+s)^{1\over 2}
\lambda (1,s)],_s
{({{s+1}\over {{s-z}}})^{1\over 2}}\nonumber\\
& & +{1\over {\pi{\sqrt {1-z}}}}\int _{-z} ^1dr[(1+r)^{1\over 2}\lambda 
(r,1)],_r
{({{r+1}
\over {{r+z}}})^{1\over 2}} \eea
\bea 
A(z)&=&{1\over {\pi\sqrt{1+z}}}\int _z ^1ds[(1+s)^{1\over 2}\psi (1,s)],_s
{({{s+1}\over {{s-z}}})^{1\over 2}}\nonumber\\
& & +{1\over {\pi{\sqrt {1-z}}}}\int _{-z} ^1dr[(1+r)^{1\over 2}\psi(r,1)],_r
{({{r+1}
\over {{r+z}}})^{1\over 2}} \eea
\be
B(z)={1\over 2 }(A(z)^2+\kappa (z)^2) -1 \ee
We can now read off the line element in the asymptotic limit of the
metric
\be
ds^2=-\xi ^{B(z)}d\xi ^2+\xi ^{B(z)}dz^2+\xi ^{\kappa (z)+A(z)}(dx^2)^2
+\xi ^{\kappa (z)-A(z)}(dx^3)^2 \ee
If we define the cosmic time $t=\xi ^{{{(B(z)+2)}\over 2}}$, then the Kasner
exponents can be read off directly to be \cite{fkv,bv}
\be p_1={{B\over {B+2}}}, ~~p_2={{\kappa +A}\over {B+2}}, ~~
p_3={{\kappa-A}\over {B+2}} \ee
The exponents satisfy the constraints,
\be
p_1^2+p_2^2+p_3^2=1,~~~\phi =(p_1+p_2+p_3-1){\rm log}~\xi \ee
There is a region in the parameter space such that all the exponents are
allowed to take negativing values and at the same time they can satisfy
the constraints as has been discussed for the four dimensional case \cite{fkv}. 
For such a case, there is a curvature singularity, and this singularity
has the characteristics of a cosmological singularity, which is very
interesting in the PBB scenario.

Now we would like to examine, the singularity structure for the 
new backgrounds generated through specific $O(d,d)$
transformations  introduced earlier. 
Note that the shifted dilaton, $\bar \phi$ remains
invariant under $O(d,d)$ transformation. The two elements
of the vielbein, $E$ are given by
\be
E_1={\rm exp}~[{1\over 2}(\lambda +\psi)]=\xi ^{{1\over 2}(\kappa +A)},~~
E_2={\rm exp}~[{1\over 2}(\lambda -\psi)]=\xi ^{{1\over 2}(\kappa -A)} \ee
Now the transformed metric becomes
\be
 G' _{22}= {{E_1}\over {(a^2+c^2E_1E_2)}},~~ G' _{33}=
{{E_2}\over {(a^2+c^2E_1E_2)}} \ee
The explicit forms of the elements, $E_1 ~{\rm and} ~~E_2$ are given above.
Thus, near the singularity, when $E_1,E_2$ go to large values, the elements
of the transformed metric remain finite. Similarly when $E_1, E_2$ tend
to zero, the structure of ${\tilde G}_{ij}$ tells us that its matrix
elements also tend to zero as $E_1$ and $E_2$ respectively. Of course, the
diagonal structure is maintained. Now turning to the dilaton in the
transformed theory, 
\be
 \phi ' =\phi - {1\over 2}{\rm log}~({{{\rm det}G}\over {{\rm det}{
G '}}})= \phi -{\rm log}~(a^2+c^2e^{\lambda})  \ee
Note, however, that in the two extreme limits of $E_1, E_2$ either tending
to zero or to infinity, $B_{ij}$ tends to a constant as is evident from
(\ref{eqb}). 

Let us proceed to derive the monodromy matrix for the above $D=4$ case which
in turn will provide connections with integrable system. The relation 
(\ref{linears}) suggests that $\hat V$ is a $4\times 4$ matrix for our case.
Moreover, the requirement of factorization of the monodromy matrix, alluded to
earlier, implies that ${\hat V }(t)$ will have a simple pole structure in
the spectral parameter $t$. In turn, it ensures that ${\cal M}(w)$
also has only 
simple poles. Now one can suitably modify the steps indicated in \cite{nico1}
to construct the $\hat V$ matrix and eventually the monodromy matrix,
which has the form 
\bea
{\cal M}(w)=\pmatrix{{{w_1-w}\over {{w_1+w}}} &0 & 0 & 0 \cr
             0 & {{w_3-w}\over {{w_3+w}}} & 0 & 0 \cr
             0 & 0 & {{w_1+w}\over {{w_1-w}}} & 0 \cr
             0 & 0 & 0 &  {{w_3+w}\over {{w_3-w}} } \cr} \eea
This form of the monodromy matrix satisfies all the expected properties.
Note from (\ref{fourd}) that the shifted dilaton, in region $IV$, can be
written as sum of two functions: one depends on $u$ and other on $v$, which
is similar to the decomposition employed in \cite{nico1}. As a consistency
check, we have considered the case when $\psi=0$, that appears in metric 
$G_{ij}$ and one could recover the known results as discussed in \cite{nico1}.
Note that, our discussion can be generalized to the case when $G$ is a 
$d\times d$ matrix as was considered in \cite{bv}; however, the number of 
poles for the $2d\times 2d$ dimensional $\hat V$ matrix will be more.

To summarize, in this note, we have presented a coset space reformulation 
of the string effective action to study the collision of plane-fronted
waves corresponding to massless states of closed string. In this process,
we have expressed the action in a form which makes it suitable to
study the
integrability properties of the theory. 
We have shown that the techniques employed in the study of integrable
systems  in
the presence of gravity can be generalized to the case of plane
wave scatterings in string theory. The monodromy matrix is derived explicitly
for the $D=4$ case. We have presented an illustrative
example, in the case of four space-time dimensions, of how one can generate
new background configurations from known solutions in the collision process.
We have exploited the properties of the coset space reformulation to generate
a space-time dependent $B$-field keeping the transformed metric
diagonal. We have only used a special type of $O(2,2) ~{\rm and}~ O(2)\times
O(2)$ transformation to generate new backgrounds. One can employ more
general form, such as space-time dependent $O(2) \times O(2)$ function $h(x)$,
for generating a wide variety of background configurations.
The approach presented here can be applied to more general situations while
considering plane wave scatterings in the context of string theory. 

\noindent {\bf Acknowledgement:} One of us (JM) would like to thank
G. Veneziano for discussions on  their work \cite{bv}. We would like to
thank H.  Nicolai for discussions and suggestions.
One of us (JM)
acknowledges gracious hospitality of   H. Nicolai  and 
the Albert Einstein Institute. This work is supported in part
by US DOE Grant No. DE-FG 02-91ER40685.

\end{document}